\pdfoutput=1

\documentclass{jfm}
\usepackage{amsmath}
\usepackage{epsfig,psfrag}
\usepackage{amsfonts}
\usepackage{mathrsfs}
\usepackage{amssymb}
\usepackage{bm}
\usepackage{url}
\usepackage{color}
\usepackage{natbib}
\usepackage{mdwlist}
\usepackage{xr}

\begin{document}
% RRK %%%%%%%%%%%%%%%%%%%%%%%%%%%%%%%%%%%%%%%%%%%%%%%%%%

\newcommand\beq{\begin{equation}}
\newcommand\eeq{\end{equation}}
\newcommand\beqa{\begin{eqnarray}}
\newcommand\eeqa{\end{eqnarray}}
\newcommand{\Sy}{{\cal S}}
\newcommand{\U}{{\cal U}}
\newcommand{\K}{{\cal K}}
\newcommand{\T}{\mathsfi {T}} % Schmidt number
\newcommand{\bx}{\mathbf{x}}
\newcommand{\by}{\mathbf{y}}
\newcommand{\bz}{\mathbf{z}}
\newcommand{\hx}{\mathbf{ \hat{x}}}
\newcommand{\hy}{\mathbf{ \hat{y}}}
\newcommand{\hz}{\mathbf{ \hat{z}}}
\newcommand{\eighth}{\footnotesize {1 \over 8}}
\newcommand{\half}{\footnotesize   {1 \over 2}}

%%%%%%%%%%%%%%%%%%%%%%%%%%%%%%%%%%%%%%%%%%%%%%%%%%%%%%%%%%%%%%%%%%%%%%%%

\def\half{\frac{1}{2}}
\def\quart{\frac{1}{4}}
\def\d{{\rm d}}
\def\eps{\epsilon}

\title{Sustaining processes from recurrent flows in body-forced turbulence}
\vspace{10mm}

\author{Dan Lucas\aff{1}\corresp{\email{dl549@cam.ac.uk}} \& Rich Kerswell\aff{2}}
\affiliation{\aff{1}{DAMTP, University of Cambridge, Cambridge, CB3 0WA, UK.}
\aff{2}{School of Mathematics, University of Bristol, University Walk, Bristol, BS8 1TW, UK.}}
\maketitle
\begin{abstract}

By extracting unstable invariant solutions directly from body-forced three-dimensional turbulence,
we study the dynamical processes at play when the forcing  is large scale and either unidirectional in the momentum or the vorticity equations. In the former case, the dynamical processes familiar from recent work on linearly-stable shear flows  - variously called the Self-Sustaining Process \citep{Waleffe:1997ia} or Vortex-Wave Interaction \citep{Hall:1991eq, HallSherwin2010} -  are important even when the base flow is linearly unstable. In the latter case, where  the forcing drives Taylor-Green vortices, a number of mechanisms are observed from the various types of periodic orbits isolated. In particular, two different transient growth mechanisms are discussed to explain the more complex states found.

\end{abstract}

%\begin{keywords}
%Turbulence theory, nonlinear dynamical systems, chaos.
%\end{keywords}
%%%%%%%%%%%%%%%%%%%%%%%%%%%%%%%%%%%%%%%%%%%%%%%%%%%%%%%%%%%%%%%%%%%%%%%%

\section{Introduction}
The chaotic and multiscale nature of turbulent flows continues to challenge the understanding of fluid dynamicists. One route to penetrating such a mathematically demanding system is to view a turbulent flow as a trajectory through a high dimensional phase space \citep{Hopf48}. This trajectory then passes near the vicinity of unstable solutions and accumulates characteristics of these states which can be extracted, via numerical computation, for thorough investigation. Such an approach was pioneered for wall-bounded shear flows \citep{Kawahara:2001ft,vanVeen:2006fm,Viswanath:2007wc,2010PhST..142a4007C,Kreilos:2012bd,Kawahara:2012iu,2015arXiv150405825W} and 2D body-forced Kolmogorov flow \citep{Chandler:2013fi,Lucas:2015gt}.

These unstable states, particularly the  time-dependent recurrent flows or UPOs (unstable periodic orbits), contain valuable dynamical information which may provide some predictability to the inherent complexity of the turbulence. This has been addressed through the application of `periodic orbit theory' \citep{Cvitanovic:1992ih} whereby the statistics of individual UPOs are combined weighted by  their relative stability in an attempt to rationalize  the turbulent statistics. Recent work \citep{Lan:2004ch,Chandler:2013fi,Lucas:2015gt} has assessed the effectiveness of this strategy and discussed computational nuances.  Rather than attempt this for three-dimensional Kolmogorov flow, we use the extracted recurrent flows as a guide for what complicated and interlinked physical processes may underpin the turbulence. A very successful example of such an approach  is plane Couette flow, where a nearly-recurrent flow was found \citep{Hamilton:1995kt} which suggested the `self-sustaining process' \citep{Waleffe:1997ia}  subsequently found manifested in a UPO extracted from  turbulent data \citep{Kawahara:2001ft}.
We consider three-dimensional  flow since this permits a greater variety of forcing than our previous 2D study   \citep{Lucas:2015gt}, in particular, enabling a  comparison between  shear-driven (closely associated with Couette flow) and vortically-driven turbulence. The vortical case has been examined previously for sustaining processes \citep{Goto2008,Goto2012,2014FlDyR..46f1413Y,Goto:2016wr} and so is a natural choice to tackle with a recurrent flow analysis. 
%To this end we discuss two possible mechanisms potentially at work in the recurrent flows we find %\citep{ANTKOWIAK:2007kz,Gau:2014cy}.
%%%%%%%%%%%%%%%%%%%%%%%%%%%%%%%%%%%%%%%%%%%%%%%%%%%%%%%%%%%%%%%%%%%%%%%%

\section{Formulation}
We consider the three dimensional, body-forced, incompressible Navier-Stokes equations
\begin{align}
\frac{\partial \bm u}{\partial t} + \bm u\cdot\nabla\bm u +\nabla p 
&= \frac{1}{Re} \Delta \bm u + \bm f \label{NSu},\\ 
\nabla\cdot \bm u &=0
\end{align}
where the Reynolds number $Re:= \sqrt{\chi}/\nu (L_f/2\pi)^{3/2}$
with $L_f$ the forcing length scale, $\chi$ the forcing amplitude and $\nu$ the kinematic viscosity.
Two forms of forcing were considered - the first, $\bm f_u=\sin 2y \sin 2z \,\hx$, driving a unidirectional basic flow ${\bm u}_{lam}:= \eighth Re \sin 2y \sin 2z \, \hx$ and the second, $\bm f_\omega:=\sin 2y \cos 2z \, \hy - \cos 2y \sin 2z \,\hz$, driving a unidirectional vorticity ${\bm \omega}_{lam}:=\half Re \sin 2y \sin 2z \, \hx$.
The equations were solved over the periodic cube $[0,2\pi]^3$ using a fully dealiased (two-thirds rule) pseudospectral method implemented in CUDA to run on a GPU card (see \cite{Lucas:2014ew} for details). A resolution of $128^3$ was used throughout which allowed the code to run locally (and efficiently) on a single GPU card (a M2090 6GB Tesla card) and the flow field was initiated with uniform amplitude but randomised phase Fourier modes in the range $2.5 \leq |\bm k | \leq 9.5$ such that the total enstrophy $\langle |\bm \omega |^2\rangle_V := 1/(2\pi)^3 \int \!\!\int\!\! \int |\bm \omega |^2 \, dx\,dy\,dz = 1$. To undercover recurrent processes embedded in the turbulent attractor, a very long DNS data set was produced and episodes where the flow appeared to almost repeat are identified on the fly. Flow fields from these episodes were then used to search for exactly recurring flow solutions using a Newton-GMRES-Hookstep algorithm \citep{Viswanath:2007wc,Chandler:2013fi, Lucas:2015gt} which took months of cpu time even with GPU acceleration. Every recurring solution found was examined for its underlying dynamical processes.

%%%%%%%%%%%%%%%%%%%%%%%%%%%%%%%%%%%%%%%%%%%%%%%%%%%%%%%%%%%%%%%%%%%%%%%%

\section{Results}

% --------------------CASE A------------------------------------------------------------------------------------------------------------------

%
% Fig 1
%
\begin{figure}
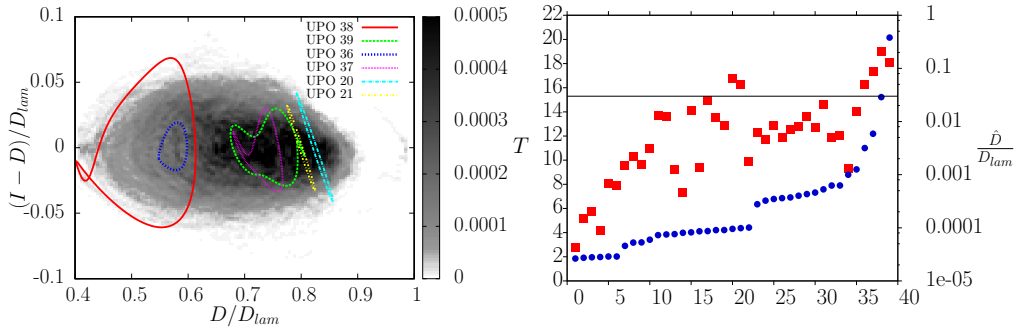

\begin{center}
\hspace{-5mm}\scalebox{0.46}{\input{DI_CASEA}}\hspace{1.2mm}
\scalebox{0.49}{\input{UPO_20_D}}
\caption{\label{fig:DI} Left: projection of dynamics onto the normalised $D$-$(I-D)$ plane, where $I=\langle \bm u \cdot \bm{f}_u\rangle_V$ is the energy input rate and  $D= \langle |\nabla\bm u |^2\rangle_V/Re$ is the dissipation rate: grey colours represent p.d.f. of the DNS with forcing $\bm{f}_u$ and  lines represent converged unstable periodic orbits (. Right:  distribution of period $T$ (blue circles) and normalised dissipation range $\hat D/D_{lam}$ (red squares)  over the UPOs found.  }
\end{center}
\end{figure}

\subsection{Case $\bm{f}_u$}

At $Re=20$  (and  $R_{thres}=0.8$ - see \cite{Lucas:2015gt}\,), a DNS data set of duration $T=6\times10^4$ (in units of $\sqrt{L_f/2\pi \chi}$)  yielded a total of 233 recurrent episodes which allowed 64 recurrent flows - 39 UPOs, 19 travelling waves (TWs) and  6 equilibria (EQs)  - to be converged. The UPOs found are documented in figure \ref{fig:DI} (and the supplemental material).   UPOs of longer period are more difficult to isolate because they tend to be more strongly unstable (both larger unstable growth rates and more unstable directions) than shorter period UPOs and as a result near-recurrent episodes in the DNS are rarer {\em and} subsequent convergence more challenging. However,  longer period orbits tend to span the turbulent attractor better (as measured by $\hat D/D_{lam}:=(\max \,D(t)-\min \,D(t) \,)/D_{lam}$ where $D= \langle |\nabla\bm u |^2\rangle_V/Re$ is the total dissipation) and should be more representative of the dynamical processes at play so we concentrate on these subsequently. A threshold of ${\hat D}=0.03D_{lam}$ (see figure \ref{fig:DI} (right)\,) picks out six UPOs - 20, 21, 36, 37, 38, 39 - which are compared with a pdf of the DNS in  figure \ref{fig:DI} (left).  

\begin{figure}
\begin{center}
%\begin{tabular}{cc}
%\hspace{-15mm} 
\parbox[b]{0.5\textwidth}{ \scalebox{0.33}{\input{fig_03}}
 \vspace{10mm}}
\hspace{3mm} 
\includegraphics[width=0.45\textwidth]{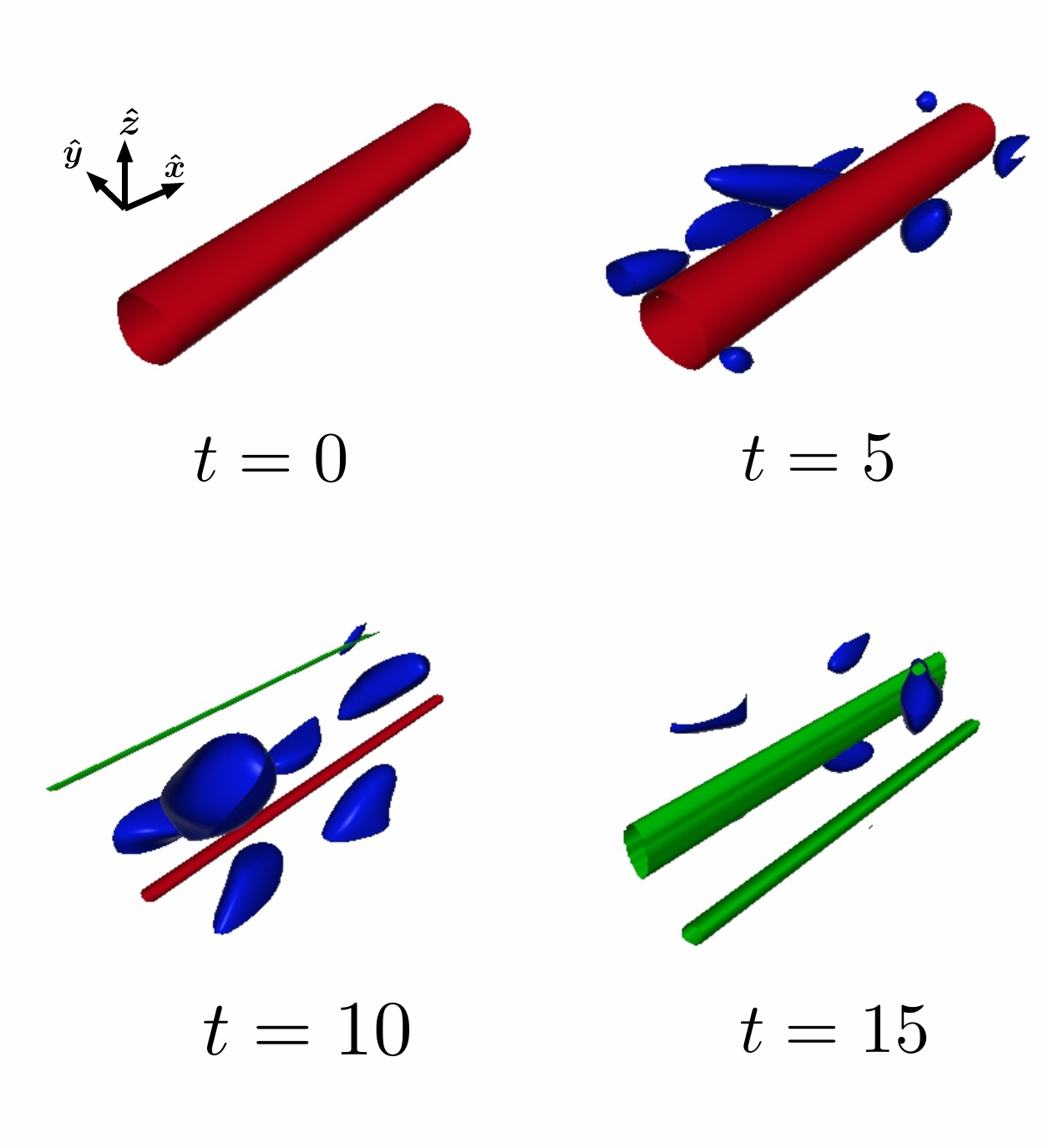}
%\end{tabular}
\caption{\label{fig:SSP} Plots showing the SSP decomposition for UPO 39. Left: time series for the energies $E_{streak}$ (red), $E_{roll}$ (green) and $E_{wave}$ (blue). Right: isosurfaces for those velocity components contributing to those energies; $\bar{u}_p=0.75$ red, $\sqrt{\bar{v}_p^2+\bar{w}_p^2}=0.13$ green and $|{\bm u}_p-\bar{\bm u}_p|=1$ for a sub-domain chosen to illustrate the cycle of behaviour.}
\end{center}
\end{figure}

% analysis
Of these UPOs, the final four show the characteristic features of the SSP/VWI  of wall-bounded shear flows in that the peak amplitudes of (small) streamwise rolls, streaks and (streamwise-dependent) waves are respectively staggered in time. The now-well-known loop of events \citep{Waleffe:1997ia,Hall:1991eq, HallSherwin2010} which produce this starts with the streamwise rolls  advecting the underlying shear to drive streaks which, when they are large enough, become unstable to streamwise-dependent  waves. These waves then grow by extracting energy from the streaks (which then diminish) and through wave-wave self-interaction re-energise the rolls whereupon  the process repeats. What is interesting is the SSP/VWI process is usually invoked  to explain the existence of solutions other than the basic state when the latter is linearly stable. Here, we find the same process operative even though the basic state is linearly {\em unstable}. To show this, the flow field of UPO 39  with the laminar state subtracted ($\bm u_p:={\bm u} - {\bm u_{lam}}= u_p \hx+v_p \hy+w_p \hz$) is decomposed into streaks $\overline{u}_p \hx$, rolls $ \overline{v}_p \hy+\overline{w}_p \hz$ and streamwise-dependent waves ${\bm u_p}-\overline{\bm u}_p$ with the corresponding energies being  defined as 
\begin{align}
E_{streak} = \half \langle    \overline{u}_p^2                                  \rangle, &\label{eq:SSP1}\quad
E_{roll}      = \half \langle    \overline{v}_p^2 +\overline{w}_p^2     \rangle  ,\quad
E_{wave}  = \half \langle \, ( {\bm u_p }-\overline{\bm u}_p \,)^2 \rangle 
%\label{eq:SSP3}
\end{align}
so that $E:=\half \langle {\bm u_p^2} \rangle =E_{streak}+E_{roll}+E_{wave}$ where $\overline{(\cdot)}:=\int^{2 \pi}_0 (\cdot) dx/2\pi$ and $\langle (\cdot)  \rangle:=\int \! \! \int \! \! \int (\cdot)\,dxdydz/(2\pi)^3$. Figure \ref{fig:SSP} shows time series for these energies, relative to the total energy,  and 3D isosurfaces for a subregion best exhibiting the regeneration cycle (secondary frequencies in the time series relate to individual streaks undergoing an SSP cycle in/out of phase with each other): initially large streaks (t=0) pump their energy into the wave field (t=5) which then decays pouring  energy into the rolls (t=10 and t=15) which subsequently drive the streaks.

\begin{figure}
\begin{center}
%\begin{tabular}{cc}
\hspace{-5mm} 
\parbox[b]{0.5\textwidth}{\includegraphics[width=0.5\textwidth]{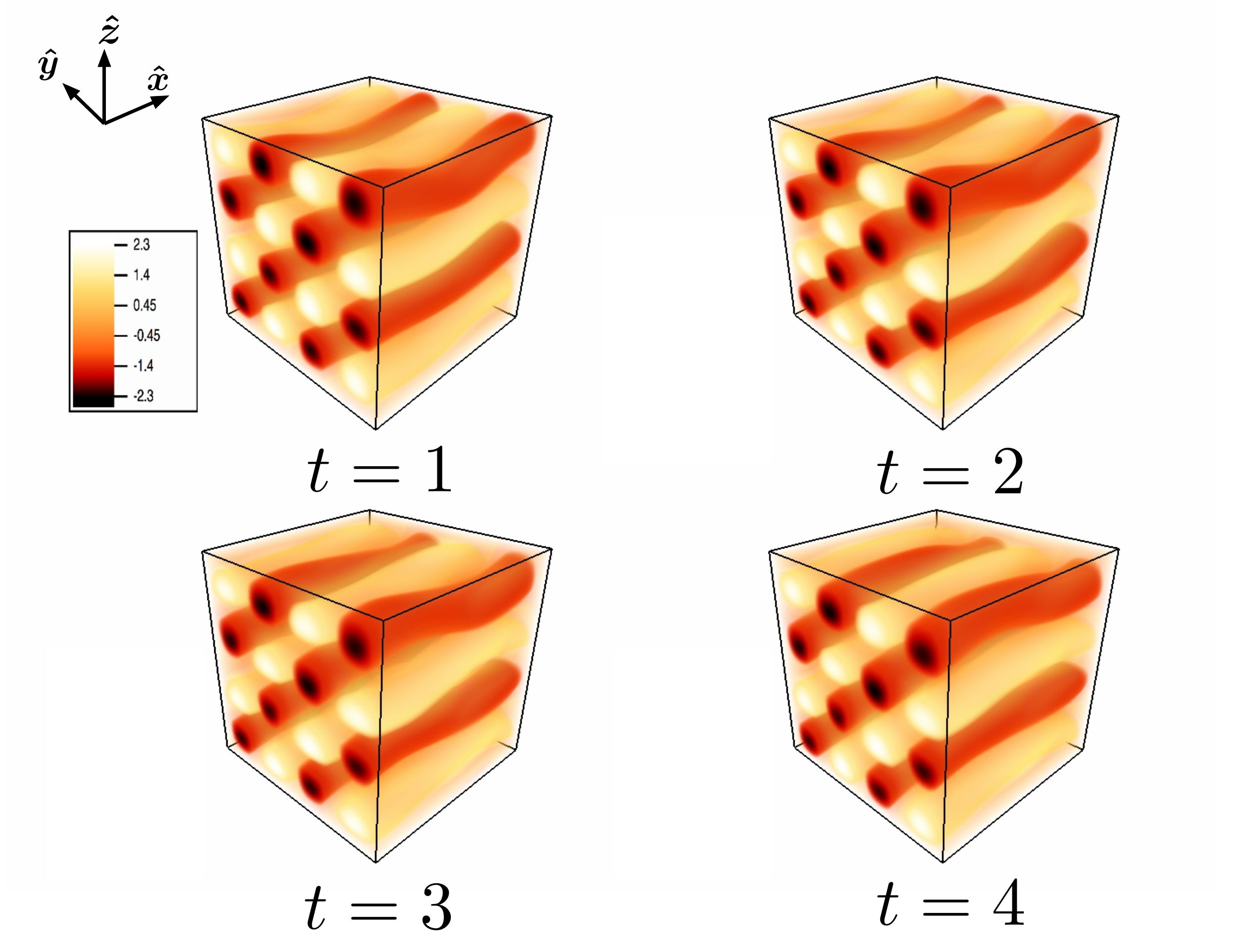}\vspace{5mm}}
% \hspace{5mm}
\scalebox{0.65}{\input{bif_plot}}
%\end{tabular}
\caption{\label{fig:UPO_20} Left: figure shows rendering of the streamwise velocity, $u\hx,$ for UPO 20 at four snapshots. This shows the propagation of long-wavelength streamsize disturbances along the base flow. Right: shows a bifurcation diagram for UPOs 20 and 38 showing, through a sequence of bifurcations (three) the connection to the base flow $u_{lam}$ (where $D/D_{lam}=1$) at $Re\approx17$ for UPO 20, but no such connections for UPO 38. }
\end{center}
\end{figure}

%  why
The key component of the SSP/VWI mechanism (in the simplest steady setting) is the presence of a neutral (small-amplitude) wave mode riding on the much larger streamwise-averaged streamwise flow $\overline{u}(y,z) \hx$.   For a linearly-stable base flow $u_{lam}(y,z)\hx$, small finite amplitude rolls are required to adjust this to make it marginally-stable (with the neutral mode so generated nonlinearly feeding back to sustain the rolls). For a linearly {\em unstable} flow, the same can clearly still happen but now the rolls can act to {\em stabilise} $u_{lam}(y,z)\hx$ by making one (or perhaps more) unstable wave  mode(s)  on $u_{lam}(y,z)\hx$ neutral on $\overline{u}(y,z)\hx$. The same thinking carries over to the more generic time-periodic situation with the rolls generating streamwise flows which are on average neutrally stable to a wave mode (or collection of wave modes). For example,  there are 12/18 unstable directions of the streak field $\overline{u}_p$ for UPOs 38/39 when the streak field is maximum and  2/14 when it is minimum. This would then create finite-amplitude states through saddle node bifurcations familiar from flows like plane Couette flow and pipe flow typically unconnected to the laminar state (e.g. see UPO 38 in figure \ref{fig:UPO_20}(right)\,).

Since $u_{lam}$ is unstable, there are also `derivative' states which arise through a continuous sequence of bifurcations off the laminar state.  UPO 20 is a typical example shown in figure \ref{fig:UPO_20} (left). Long-wavelength streamwise disturbances propagate on top of the base flow, consistent with the long-wavelength instability of Kolmogorov flow \citep{Meshalkin1961} and the streak field is small and varies very little. As a result, there is no stability change of $u(y,z) \hx$ for  UPO 20 over a cycle: 20 unstable directions are always found for $\overline{u}_p$  compared to the 32 unstable directions for $u_{lam}(y,z) \hx$. The overall conclusion of our computations at $Re=20$ is that there are SSP/VWI UPOs and `derivative' UPOs which are {\em both} dynamically important in the complex dynamics observed.

% examples

%
% Fig 2
%

%
% Fig 3
%

At $Re=30$ a time integration of $T=1.6\times 10^5$ finds 67 guesses, however, none of these can be converged to invariant solutions presumably because these are becoming more unstable. To quantify this, all six large-${\hat D}$ UPOs were continued up in $Re$ from 20 using the arc-length continuation. All continuations quickly turned back (figure \ref{fig:UPO_20} shows two such continuations) retreating to lower $Re$ with only 3 out of the 6 UPOs (UPOs 20, 21 and 38) extending beyond $Re=30$. UPO 21 reaches the furthest ($Re=42$) and by then its instability has considerably increased by two complementary measures:  a) the sum of the unstable growth rates more than quadruples and b) the number of unstable directions jumps from 20 to 93. The other UPOs show similar rapid destabilization (e.g. the dimension of the unstable manifold for UPO 37 jumps from 15 at $Re=20$ to 58 at $Re=27$ while that for UPO 20 increases from 25 at $Re=20$ to 67 at $Re=30$). These figures suggest that the dimension of the chaotic attractor very quickly increases as $Re$ increases beyond 20.

%
%
%------CASE B--------------------------------------------------------------------------------------------------------------------------
%
%

%
% Fig 4
%
\begin{figure}
\begin{center}
\hspace{-7mm}
\scalebox{0.46}{\input{DI_CASEB}}
\hspace{2mm}
\scalebox{0.48}{\input{UPO_B_D}} % replace with correct plot for case B
\caption{\label{f_omega} Left: projection of dynamics onto the normalised $D$-$(I-D)$ plane, where $I=\langle \bm u \cdot \bm{f}_u\rangle_V$ is the energy input rate and  $D= \langle |\nabla\bm u |^2\rangle_V/Re$ is the dissipation rate:: grey colours represent p.d.f. of the DNS with forcing $\bm{f}_\omega$ and  lines represent converged unstable periodic orbits. Right: distribution of period $T$ (blue circles) and normalised dissipation range $\hat D/D_{lam}$ (red squares) over the set of 15 UPOs.}
\end{center}
\end{figure}

%
% Fig 5
%
\begin{figure}
\begin{center}
\includegraphics[width=0.47\textwidth]{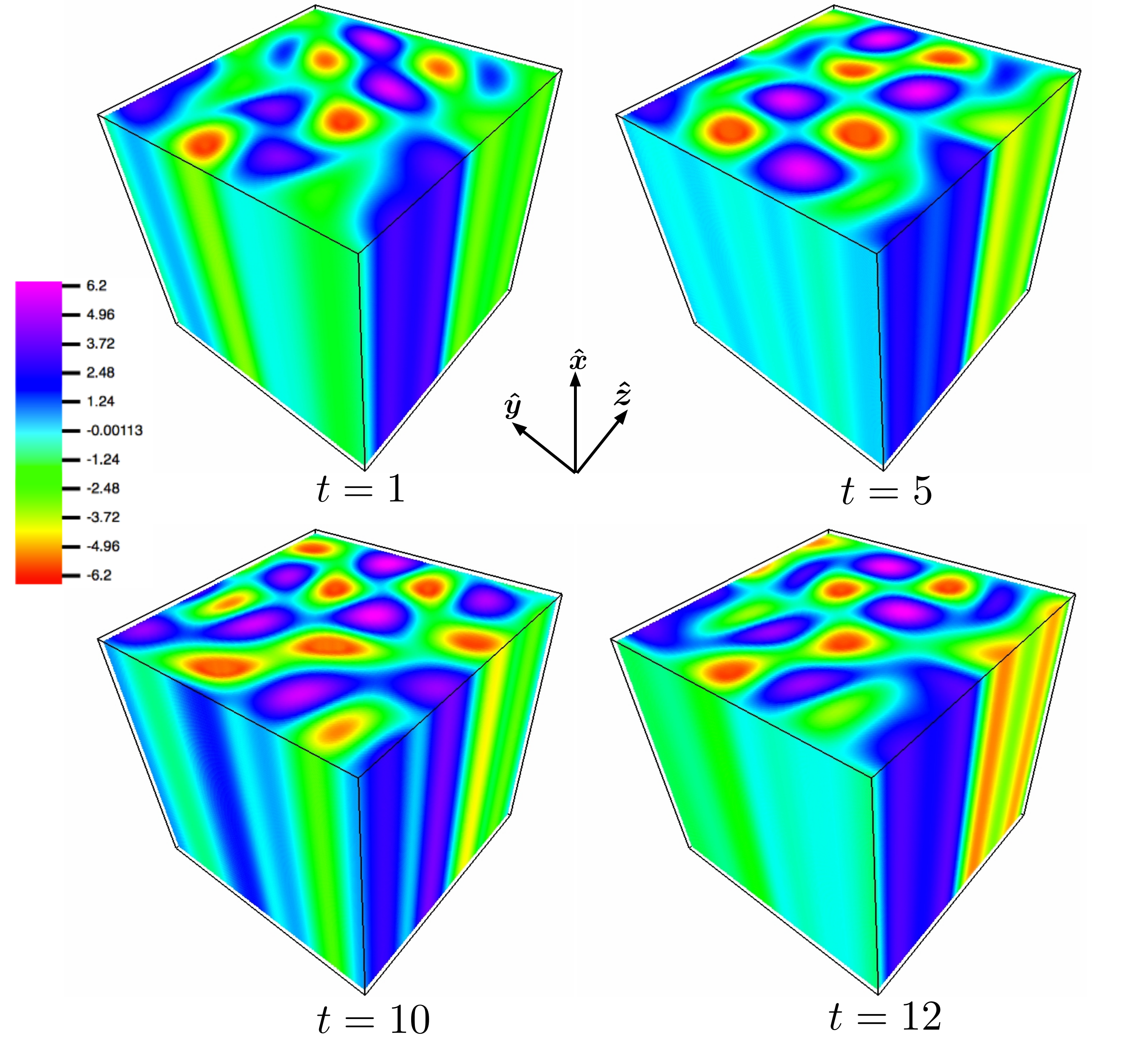}\vline
\includegraphics[width=0.45\textwidth]{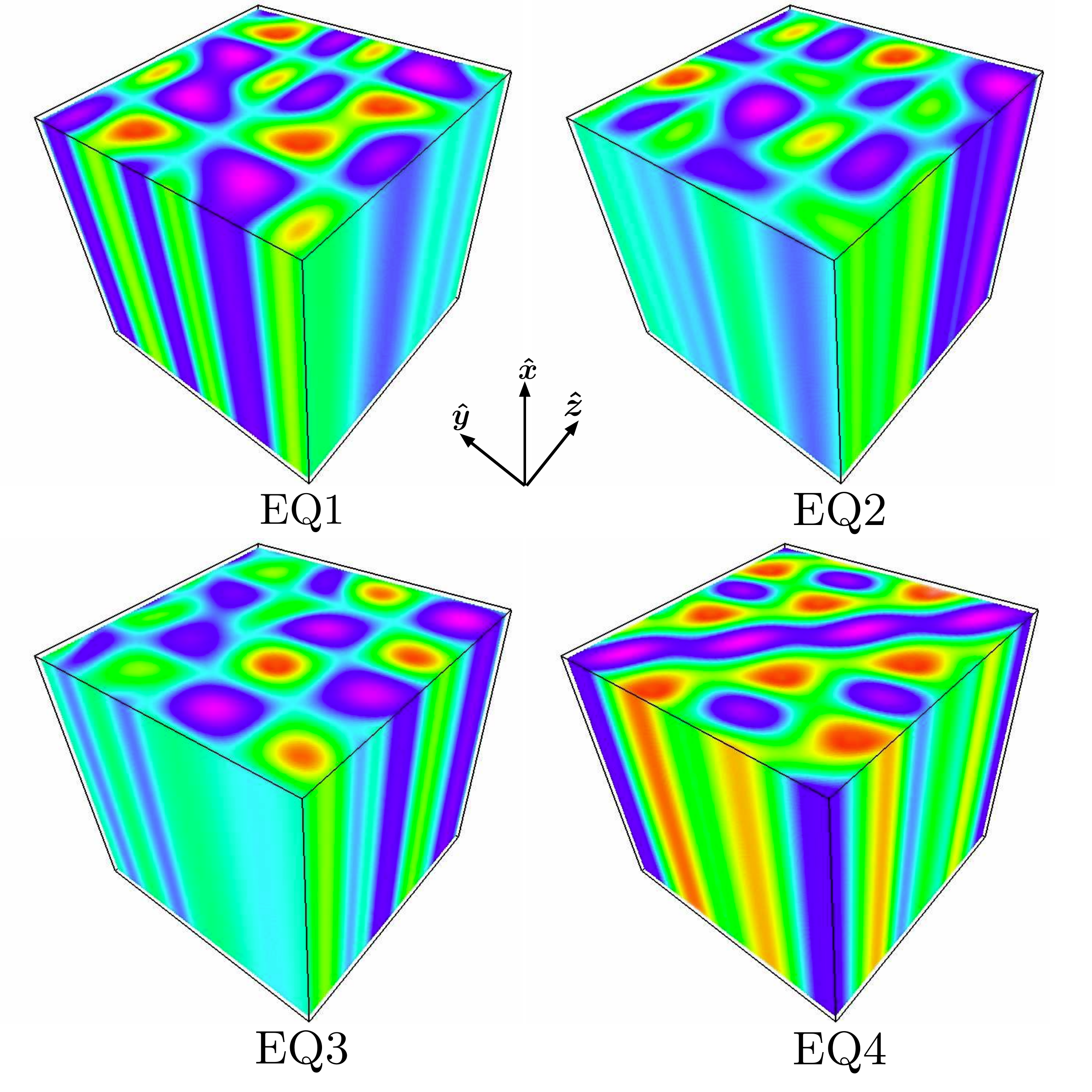}
\caption{\label{fig:UPO1_EQsii} Snapshots of isometric view of axial vorticity $\xi$ for the two-dimensional UPO 15 (left) and steady states EQ1-4, (right) in case $\bm{f}_\omega.$ }
\end{center}
\end{figure}

\subsection{Case $\bm{f}_\omega$}
For unidirectional vorticity forcing at $Re=15$, a DNS of duration  $T=10^4$ with $R_{thres}=0.8$ yielded a total of 600 recurrent episodes from which 22 unstable recurrent flows - 15 UPOs, 3 TWs and 4 EQs  -  were converged with the Newton-GMRES-hookstep algorithm. Figure \ref{f_omega} shows the projection of the six UPOs with largest $\hat{D}$ and a p.d.f. of the DNS trajectory  on the  $(I-D)/D_{lam}$ versus $D/D_{lam}$ plane. These solutions can be classified into three categories.

% type 1
The simplest solutions are 2-dimensional. As an example,  UPO 15, which has the largest $\hat{D}$, shows a classical 2D loop of vortex merger and splitting with no variations in the $x$ direction. This type of process was examined in detail in 2D Kolmogorov flow in \citep{Lucas:2015gt}.  All four equilibria are 2D states and are shown in  figure \ref{fig:UPO1_EQsii}.

\begin{figure}
\begin{center}
\includegraphics[width=\textwidth]{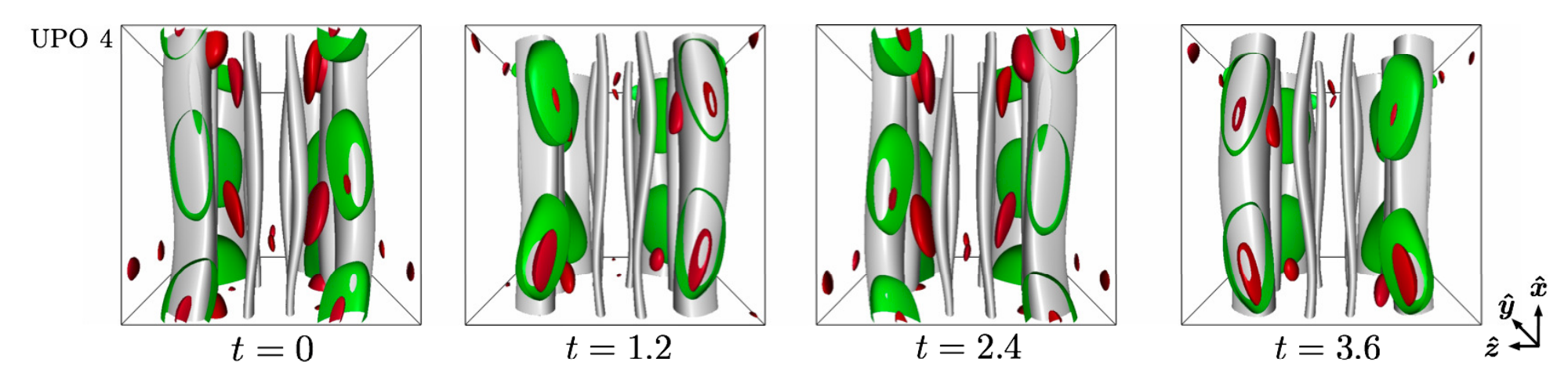}
\caption{\label{fig:UPO4ii} Isometric view of vorticity isosurfaces ($|\xi|=3$ coloured grey, $|\eta|=0.5$ coloured red and $|\zeta|=0.5$ coloured green) for UPO 4 in case $\bm{f}_\omega.$ This UPO exhibits the growth and propagation of bending modes of the basic vortex tubes.}
\end{center}
\end{figure}

% type 2
The second type of solution breaks the $x$-translational symmetry, but shows quite low amplitude time dependence in the form of vortex waves (Kelvin modes) on the basic vortex tubes.  This describes most of the low amplitude (small $\hat{D}$) UPOs with figure \ref{fig:UPO4ii} showing an example, UPO 4, which was the UPO converged most frequently (8 times) from the recurrent guesses. Here, as in the other low-amplitude UPOs, the three-dimensionality is provided by a dominant $k_z=1$ long axial wavelength in the solutions.  These first and  second types of solution are clearly `derivative' states which can be smoothly traced back, via a small number of bifurcations, to the laminar solution as $Re$ is reduced.

%
% Fig 6
%

\begin{figure}
\begin{center}
\includegraphics[width=\textwidth]{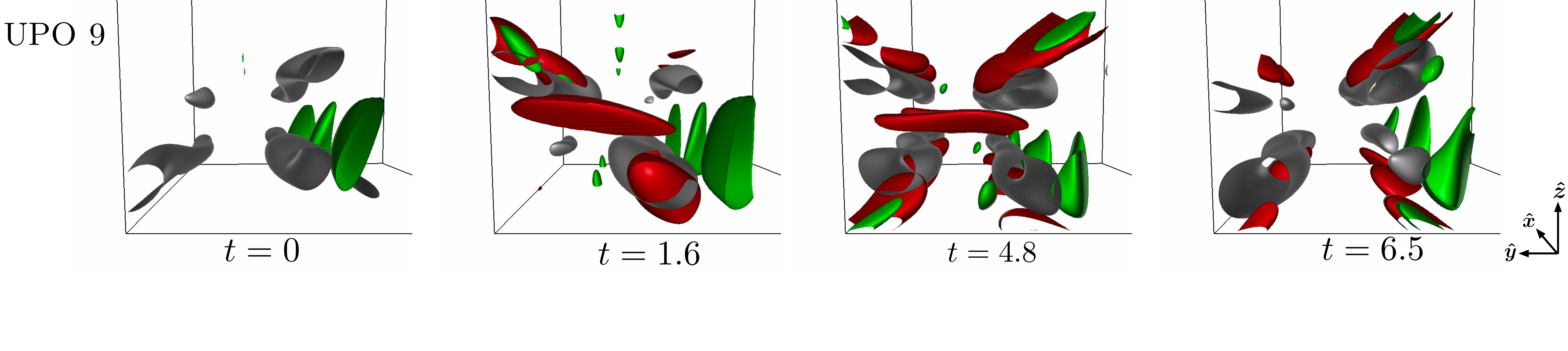}
\caption{\label{fig:UPO2ii} Isometric view of vorticity isosurfaces ($|\xi|=4$ coloured grey, $|\eta|=1.5$ coloured red and $|\zeta|=1.0$ coloured green) for UPO 9 in case $\bm{f}_\omega.$ Notice here we show a quarter of the domain to focus attention on the generation of horizontal vorticity about the stagnation point. }
\end{center}
\end{figure}

\begin{figure}
\begin{center}
\includegraphics[width=\textwidth]{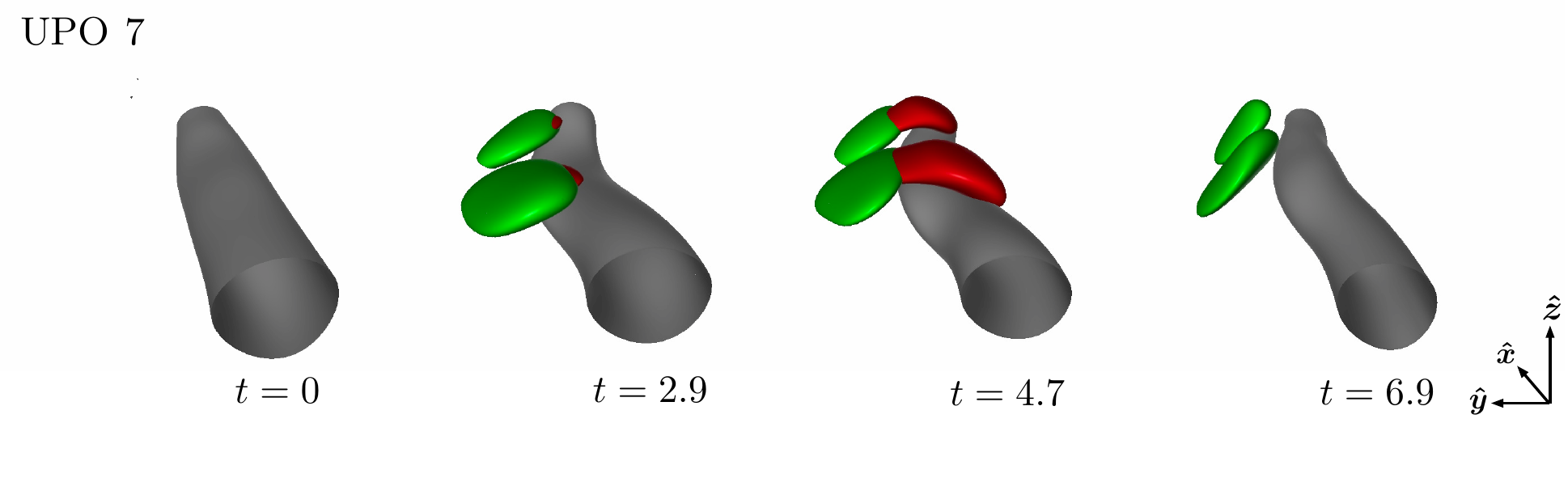}
\caption{\label{fig:UPO6ii} Isometric view of vorticity isosurfaces ($|\xi|=4.5$ coloured grey, $|\eta|=1$ coloured red and $|\zeta|=1$ coloured green) for UPO 7 in case $\bm{f}_\omega.$ Notice here we show a subdomain to focus attention on the generation of azimuthal vorticity around a vortex tube. }
\end{center}
\end{figure}

% type 3
The third type of solution has  larger $\hat{D}\,$ (excluding the outlying 2D state, UPO15) with UPOs 7 and 9 being prime examples. These UPOs exhibit some of the characteristics of the other solutions, i.e.  shift-\&-reflect and $x-$invariance symmetry breaking, but also more complex nonlinear behaviour too. This is generally manifest by larger amplitude bending of axial vortex tubes and consequently more substantial production of horizontal vorticity. The horizontal vorticity is observed principally organised at the hyperbolic stagnation points, e.g. figure \ref{fig:UPO2ii}, but sometimes as localised toroidal vortex tubes about the basic axial vorticity, e.g. figure \ref{fig:UPO6ii}. 

%
%  mechanism 1:    vorticity growth near stag. pts
%
Figure \ref{fig:UPO2ii} shows a quarter  of the domain of UPO 9 and highlights the production of stagnation point horizontal vorticity in the form of a thin vortex sheet. This occurs when the main axial vortex tubes are arranged such that adjacent vortices are  bent toward each other, thereby increasing the local shear. In this region the anti-parallel vortex lines will viscously reconnect if they approach closely enough, resulting in horizontal vorticity which is then stretched into thin sheets by the background strain. This has some similarity to the transient growth enhancement in Taylor-Green vortices reported by \cite{Gau:2014cy} where linear non-modal energy optimals show concentrated growth near the hyperbolic stagnation points. Following this production phase, the thin sheet decays due to viscosity and the axial vorticity is re-energised by the forcing. Note that while this is observed in one vortex `cell', other vortex tubes are meanwhile undergoing similar, but distinct, interactions out of phase with the one presented here. This makes quantitative analysis of precise isolated mechanisms a challenge and indicates that precise periodicity of invariant solutions may be quite rare. 

%
% Fig 7
%

%
% mechanism 2:     Antkowiak & Brancher/Goto
%
Figure \ref{fig:UPO6ii} shows a close up of an individual vortex tube in UPO 7 as it goes through another type of periodic cycle. The initially straight vortex tube gives rise to a growing patch of azimuthal vorticity (see $t=2.9$ and $t=4.7$) before it dies away (at $t=6.9$). This evolution is reminiscent of the  `anti-lift-up' mechanism found in \cite{ANTKOWIAK:2007kz}. Here  azimuthal streaks are found to be the (linear) transient energy growth optimal for vortex tubes, evolving into much larger amplitude azimuthal vortices (rolls) as opposed  to the generic lift-up mechanism in unidirectional shear flow of  streamwise {\em rolls} growing into much larger streamwise {\em streaks} \citep{Farrell1988}. 

The existence of this `new'  transient growth mechanism could  suggest the existence of  an alternative type of self-sustaining process where the roles of the rolls and streaks (now defined relative to a different basic state so, for example `streamwise-independent' means axisymmetric) are reversed. However, \cite{Rincon2007} argued convincingly against this possibility in the context of the Keplerian disk problem where the basic flow field is well known to be linearly stable. Here our situation is rather different since the basic flow is unstable at this $Re$ and so (as argued for the shear case above) the growing azimuthal rolls have more flexibility in manufacturing  a (non-axisymmetric) wave mode  which is neutral on average about the (now axially and radially varying) vortex to drive the (small) azimuthal streaks. The core problem which \cite{Rincon2007} identified, however, remains that the evolving finite-amplitude roll field interacts nonlinearly with itself and therefore is much more tightly constrained than the streak field in the SSP which does not self-interact due to its streamwise independence. 

Nevertheless, \cite{Goto:2016wr} (see also \cite{Goto2008, Goto2012}) do  report that `daughter' vortices are often observed in vortically-driven turbulence forming  in the large scale strain regions, perpendicular to their `parents' in anti-parallel pairs and  conjecture that this may be a driving mechanism for the turbulent cascade process. Here we see a similar picture emerge from the converged UPOs (e.g. figure \ref{fig:UPO2ii}), although we don't see the same scale separation 
and only one generation of offspring both presumably due to the low Reynolds number used here.

\begin{figure}
\begin{center}
\includegraphics[width=\textwidth]{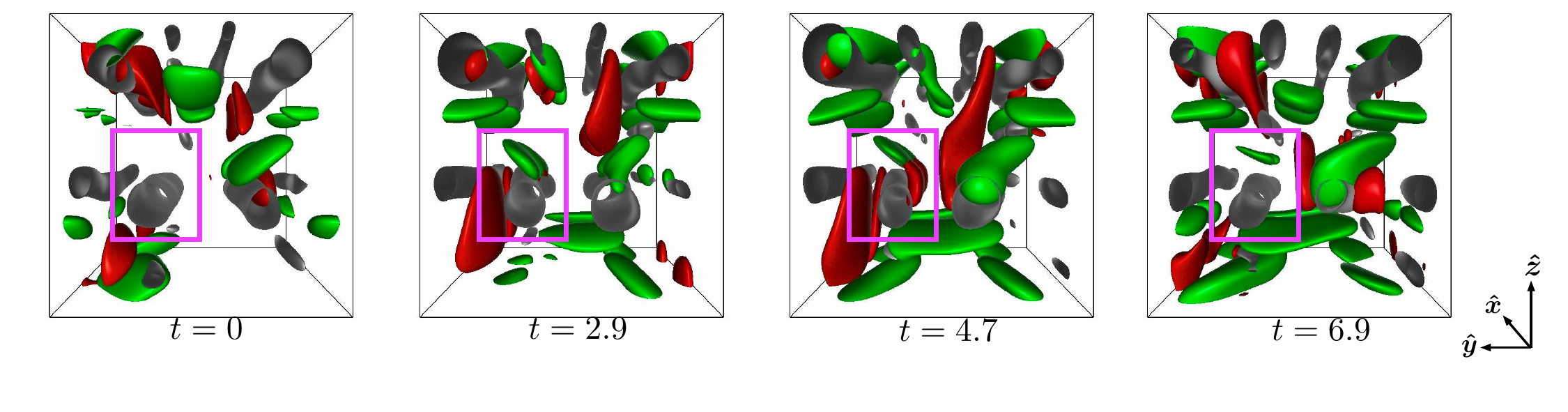}
\caption{\label{fig:UPO6ii2} Isometric view of vorticity isosurfaces ($|\xi|=4.5$ coloured grey, $|\eta|=1$ coloured red and $|\zeta|=1$ coloured green) for UPO 7 in case $\bm{f}_\omega.$ Here we show the full domain, including the subdomain of figure \ref{fig:UPO6ii} (highlighted with magenta box) in order to show the complexity of the processes occurring.}
\end{center}
\end{figure}

The full dynamics of the UPOs are typically more complex than the individual interactions we have so far highlighted. For example, figure \ref{fig:UPO6ii2} shows the full domain of UPO 7 where the coherence between different cells appears lost. Furthermore, while we have discussed two possible transient growth mechanisms postulated before \citep{Gau:2014cy,ANTKOWIAK:2007kz}, the `base-flow'  in these UPOs is \emph{not} really the low $Re$ laminar solution, but rather other averaged configurations arising from saturated instabilities such as  the 2D states shown in figure \ref{fig:UPO1_EQsii}.

%
% Fig 8
%

%
% Fig 9
%

\section{Discussion}

Unstable solutions have been extracted directly from 3D body-forced turbulence for the first time at least at modest $Re$. Two distinct forms of forcing, unidirectional in either the velocity or vorticity field, have been compared and the unstable orbits analysed to reveal possible sustaining mechanisms at work in these flows. When the forcing produces a unidirectional velocity with background shear, the UPOs found partition into two groups: those that originate from a linear instability off the base state - `derivative'  states - and those (generally of larger amplitude) that do not,  which show the characteristic features of  the SSP/VWI process \citep{Hamilton:1995kt,Waleffe:1997ia,Hall:1991eq}.  The key new realisation here is that SSP/VWI UPOs can be built upon a $u(y,z)\hx$ base flow whether it is linearly stable {\em or} unstable. Our calculations indicate that `derivative' UPOs and SSP/VWI UPOs are both dynamically important in the complex flows computed at least at $Re=20$.

The situation is more complicated when the forcing is unidirectional in the vorticity.  A number of steady and time-dependent two-dimensional states  are extracted from the 3D turbulence, indicating that at this relatively low $Re$, 2D vortex dynamics can still play a role. Beyond these states, bending modes of the main vortex tubes provide the simplest, three-dimensional behaviour. Horizontal vorticity is then observed to arise from two different transient growth mechanisms; stagnation point generation \citep{Gau:2014cy}, and local azimuthal generation \citep{ANTKOWIAK:2007kz}. The former is a feature of the global structure of the flow - driven vortices in periodic cells - and is difficult to avoid unless a single driven vortex is treated in isolation. The latter is particularly interesting in that it has recently been conjectured as the central process in the turbulent cascade \citep{Goto:2016wr} and may form a key stage in a new self-sustaining process. The existence of such a process would allow finite-amplitude states to exist without having to rely on a linear instability as an energy input. Put another way, in the absence of any such process, no other states should exist for any driven vortical state which is  linearly stable for all $Re$. This would mean no turbulence in contrast to the unidirectional shear situation where SSP/VWI exists. However, we have been unable to confirm this in the periodic flow configuration studied here where multiple vortex tubes are driven in one domain across which periodicity is assumed. This gives rise to a complicated array of linear instability modes which are either global (reaching across multiple vortex tubes) or localised 
to one tube \citep{Gau:2014cy}.  

One notable outcome of this work is the very limited success of recurrent flow analysis as $Re$ is increased.
Undoubtedly this is because the dimension of the chaotic attractor increases steeply as indicated by the rapid destabilisation of  UPOs continued up from lower $Re$.  Future work will have to concentrate on this difficulty which is clearly more pronounced in 3D than 2D \citep{Lucas:2015gt} to make searches for recurrent flows as efficient as possible.  The results presented here are the result of many months of wall time even with the substantial savings afforded by GPU-accelerated timestepping. Possible improvements may come from quotienting out the continuous symmetries \citep{Willis:2013bu}, a novel pre-processing of guesses prior to GMRES-Newton iterations \citep{Farazmand:2016hf} or a windowing method to concentrate numerical effort on some interrogation window of interest \citep{Teramura:2014gy}. 

\vspace{0.25cm}
\indent
{\em Acknowledgements}. We would like to thank Susumu Goto for helpful discussions and for sharing his submitted manuscript with us. We are also grateful for numerous free days of GPU time on ``Emerald'' (the e-Infrastructure South GPU supercomputer: \url{http://www.einfrastructuresouth.ac.uk/cfi/emerald}) and the support of EPSRC through Grant No. EP/H010017/1.

%%%%%%%%%%%%%%%%%%%%%%%%%%%%%%%%%%%%%%%%%%%%%%%%%%%%%%%%%%%%%%%%%%%%%%%%
%\newpage

\bibliography{papers}
\bibliographystyle{jfm}

\end{document}